\documentclass[preprint2]{aastex}
\newcommand{\be}{\begin{equation}}
\newcommand{\degdot}{\mbox{$. \! ^{\circ}$}} % Defines degree sign in decimal number
\newcommand{\ee}{\end{equation}}

\newcommand{\ax}{$\alpha_{\rm X}$}
\newcommand{\bx}{$\beta_{\rm X}$}

\newcommand{\plm}{$\pm$}

\newcommand{\swift}{\mbox{\it Swift}}	    % defines Chandra name
\newcommand{\meszaros}{M\'esz\'aros~}

\shorttitle{GRB050603}
\shortauthors{Grupe et al.}

\begin{document}

%\input DGrupe_clipfig.tex
%\useunitmm

\def\etal{{\it et\thinspace al.}\ }
\def\alp{{$\alpha$}\ }
\def\al2{{$\alpha^2$}\ }

%\def \charthoffset {\hspace{0.2cm}} \def \charthsep {\hspace{0.3cm}}
%\def \chartvsepcap {\vspace{0.3cm}}
%\def \chartvsep {\vspace{0.1cm}}
%\newcommand{\putchartb}[1]{\clipfig{#1}{75}{20}{7}{275}{192}}
%\newcommand{\putchartc}[1]{\clipfig{#1}{55}{33}{19}{275}{195}}
%
%
%
%\newcommand{\chartlineb}[2]{\parbox[t]{18cm}{\noindent\charthoffset\putchartb{#1}\charthsep\putchartb{#2}\chartvsep}}
%\newcommand{\chartlinec}[3]{\parbox[t]{18cm}{\noindent\charthoffset\putchartc{#1}\charthsep\putchartc{#2}\chartvsep\putchartc{#3}\chartvsep}}


\title{\swift\ Observations of GRB\,050603: An afterglow with a steep late time
decay slope
}


\author{Dirk Grupe\altaffilmark{1},\email{grupe@astro.psu.edu} 
Peter J. Brown\altaffilmark{1}, 
Jay Cummings\altaffilmark{2},
Bing Zhang\altaffilmark{3}, 
Alon Retter\altaffilmark{1},  
David N. Burrows\altaffilmark{1}, 
Patricia T. Boyd\altaffilmark{2},
Milvia Capalbi\altaffilmark{4}
Neil Gehrels\altaffilmark{2}
Stephen T. Holland\altaffilmark{2,5}, 
Peter \meszaros\altaffilmark{1,6},
John A. Nousek\altaffilmark{1}, 
Jamie A. Kennea\altaffilmark{1}, 
Paul O'Brien\altaffilmark{7},
Julian Osborne\altaffilmark{7},
Claudio Pagani\altaffilmark{1}, 
Judith L. Racusin\altaffilmark{1}, 
Peter Roming\altaffilmark{1}, 
 and Patricia Schady\altaffilmark{1,8} 
}

\altaffiltext{1}{Department of Astronomy and Astrophysics, Pennsylvania State
University, 525 Davey Lab, University Park, PA 16802} 
\altaffiltext{2}{NASA Goddard Space Flight Center, Greenbelt, MD 20771}
\altaffiltext{3}{Department of Physics, University of Nevada, Las Vegas, 
NV 89154}
\altaffiltext{4}{ASI Science Data Center, via G. Galilei, I-00044 Frascati
(Roma), Italy}
\altaffiltext{5}{Universities Space Research Association, Seabrook, MD}
\altaffiltext{6}{Department of Physics, Pennsylvania State University,
University Park, PA 16802}
\altaffiltext{7}{Department of Physics \& Astronomy, University of Leicester,
Leicester, LE1 7R, UK}
\altaffiltext{8}{Mullard Space Science Laboratory, Holmbury St. Mary, Dorking,
Surrey RH5 6NT, U.K.}


%\altaffiltext{1}{Visiting Astronomer, Cerro Tololo Inter-American Observat}


\begin{abstract}
We report the results of \swift\  observations of the Gamma Ray Burst 
GRB\,050603.
With a V magnitude V=18.2 about 10 hours after the burst 
the optical afterglow was the brightest 
so far detected by \swift\  and one of the
brightest optical afterglows ever seen. 
The Burst Alert Telescope (BAT) light curves
show three fast-rise-exponential-decay spikes with $T_{90}$=12s and a fluence
of 7.6$\times 10^{-6}$ ergs cm$^{-2}$ in the 15-150 keV band. With an 
$E_{\rm \gamma, iso} = 1.26 \times 10^{54}$ ergs it was also one of the most energetic
bursts of all times.  
The \swift\ spacecraft began observing
of the afterglow with the narrow-field instruments 
 about 10 hours after the detection of the burst. 
The burst was bright enough to be detected by the \swift\  UV/Optical
telescope (UVOT) for almost 3 days and by the X-ray Telescope (XRT) for a week
after the burst.
The X-ray light
curve  shows a rapidly fading afterglow with a decay index 
$\alpha$=1.76$^{+0.15}_{-0.07}$. 
The X-ray energy spectral index was $\beta_{\rm
X}$=0.71\plm0.10 with the column density in agreement with the Galactic
value. The spectral analysis does not show an obvious change in the X-ray
spectral slope over time. 
The optical UVOT light curve decays with a slope of $\alpha$=1.8\plm0.2.
 The steepness and the similarity of the
optical and X-ray decay rates suggest that the afterglow was observed after the
jet break. We estimate a jet opening angle of about 1-2$^{\circ}$.
\end{abstract}


%DNB changes in bold

\section{Introduction}

With an isotropic equivalent energy release on the order of
 $10^{52} - 10^{54}$ ergs, 
Gamma Ray Bursts (GRBs) are  among the most energetic events in
the Universe besides
% the actual energies of GRBs, once you correct for jet angles, seem
% to be more similar to SNe energies
the Big Bang. GRBs can be separated into two classes: short and long bursts. 
Long bursts, with durations longer than 2 seconds 
\citep[e.g.][]{chryssa93} are associated with the collapse of a very massive star and the
formation of a  black hole \citep{woosley93}.
Short bursts are thought to be the
result of a neutron star (NS) - NS or NS - black hole merger
\citep[e.g][]{eichler89, paczynski91}.
The leading 
theoretical model for GRBs and their afterglows is  the fireball model
\citep[see ][ and references within]{meszaros97, sari98, zhang04}
in which the GRB is produced by internal shocks in a 
relativistic fireball, while the afterglow is produced in external shocks
which are 
created when the fireball encounters the ambient external medium.
 
The multi-wavelength mission {\it Swift} \citep{gehrels04} was
launched on 2004-November-20 in order to hunt for GRBs.
 It is  in low-Earth orbit at an altitude of 600 kms.
The \swift\ observatory is equipped with three telescopes: a) the Burst Alert
Telescope \citep[BAT, ][]{barthelmy05}, b) the X-Ray Telescope 
\citep[XRT, ][]{burrows05}, and c) the UV-Optical
Telescope \citep[UVOT, ][]{roming05}. The BAT is a coded mask experiment and operates in the
15-350 keV energy range. The XRT detector is a copy of the MOS CCDs used 
on-board XMM \citep{tur01}. It operates between 0.3-10 keV in three observing
modes, Photon Counting (PC) which is equivalent to the full-frame mode on XMM,
Windowed Timing (WT), and Low-Rate Photo-diode mode (LrPD) which is only used for
extremely bright sources \citep{hill04}.
The UVOT is a sister instrument to XMM's Optical Monitor
\citep[OM, ][]{mason01} and includes a similar set of filters to the OM
\citep{mason01, roming05}.

As reported by \citet{nousek05}, 
\swift\ has observed a general behavior of GRB afterglow X-ray
light curves: a fast decay with slope $\alpha_1$ in the first hundred seconds  
is followed by a much
shallower decay slope $\alpha_2$. This continues over a span of several thousands
of seconds after the burst and is followed by a steeper decay slope $\alpha_3$
\citep[see also ][]{bing05}. The mean decay slopes of the 27 afterglows
discussed in \citet{nousek05} are:
$\alpha_1$=3.38\plm1.27, $\alpha_2$=0.76\plm0.34, and
$\alpha_3$=1.33\plm0.32.

In this paper we report the \swift\ observations of GRB\,050603.  
The paper % and  
is organized as follows: In \S\ref{observe} we describe the
observations and the data reduction and analysis, in \S\ref{results} we
present the results which are then discussed in \S\ref{discuss}. Throughout
the paper decay and energy spectral indices $\alpha$ and $\beta$ 
are defined as $F_{\nu}(t,\nu)\propto
(t-t_0)^{-\alpha}\nu^{-\beta}$ with $t_0$ the trigger time of the burst. 
Luminosities are calculated assuming a $\Lambda$CDM
cosmology with $\Omega_{\rm M}$=0.27, $\Omega_{\Lambda}$=0.73 and a Hubble
constant of $H_0$=71 km s$^{-1}$ Mpc$^{-1}$ using the luminosity distances
given by \citet{hogg99}. All errors are 1$\sigma$ unless stated otherwise.

\section{\label{observe} Observations and data reduction}

GRB\,050603 was detected by the BAT on 2005-June-03 at 06:29:05 UT
\citep{retter05} with the trigger ID 131560.
Due to engineering tests of the \swift\ satellite, the
narrow-field instruments, UVOT and XRT, were unable to observe the GRB 
afterglow until about 9.5 and 11 hours after the burst, respectively.

The UVOT began observing at 15:42:59 UT \citep{brown05a}. 
The observations were made in the 
V filter with exposure time ratios of 1:8:1 per observing window so that the 
entire observation is not ruined if there is high background from the earth 
limb at the beginning or end of the observation. 
 Only the middle exposures were  used, to 
 eliminate the problems of high background. The NASA \swift\ Data
Center (SDC)-processed sky images 
 were astrometrically aligned with respect to an image of the field from the 
 Digitized Sky Survey.  Aperture photometry was performed using a 6$''$ source 
 aperture and a background annulus with inner and outer radii of 14$''$ and
 30$''$ 
 respectively.  Once the afterglow in individual observations had fallen below 
 3$\sigma$ above background they were co-added to bring the S/N ratio above 
 a 3$\sigma$ detection.

The GRB\,050603 observations by the UVOT
also revealed an infrequent problem in which the 
images do not contain events from the whole exposure time indicated in the 
header.  Extension 4 of Sequence 00131560001 reported an exposure time of 1500s.
However, a comparison of the count rate measured by the instrument and the counts 
 in the image indicated that the image contained only 110 seconds of data.  
 Thus only 7.3\% of the data were actually recorded. This resulted in an 
 erroneous magnitude being reported in Brown et al. (2005a), as pointed out by 
 Berger (2005).    We have corrected this by comparing 
 the count rate as measured by the UVOT with the counts in the actual image 
 to calculate the exposure time.  When the header and calculated exposure times 
 differed by more than 5\%, the exposure keyword was changed to the 
 calculated value. The software bug causing this problem  was 
fixed on
 2005-September-14 and data taken before that date will be checked and corrected in the
 archive.

The XRT started to take data at 17:19:27 UT
\citep{racusin05}. GRB050603 was observed over a period of about two weeks for
a total of
about 220 ks. The detailed observation log is listed in Table~\ref{obs_log}.
All observations were performed in PC mode.
The XRT data were reduced by the {\it xrtpipeline} software version 0.9.9. 
Source photons were selected by {\it XSELECT} in a circular region with a radius
of r=47$''$ and the background photons in a circular region close by with a
radius r=96$''$. For the spectral data events with grades 0-12 were
selected with {\it XSELECT}.
The spectral data
were re-binned by {\it grppha} 3.0.0 having at least a minimum of 
20 photons per bin. The spectra were 
analyzed by
{\it XSPEC} version 11.3.2. The auxiliary response files were created by {\it
xrtmkarf} and the standard response matrix swxpc0to12\_20010101v007.rmf was used.

Background-subtracted X-ray light curves in the 0.3-10.0 keV energy range
were constructed using the ESO Munich
Image Data Analysis Software {\it MIDAS} (version 04Sep). The binning was dynamically 
performed. At the beginning of the observations the binning was set to 50
photons per bin while at later times it was reduced to 10 photons per bin as listed in  
Table\,\ref{obs_log}. 
Also at later times the source extraction radius was reduced to 10 pixels
(corresponding to r=23.6$''$) in order to avoid confusion by the background.
The X-ray light curve was fitted by power law and broken power law 
 models in {\it XSPEC}. 
The count rates were converted into an unabsorbed flux by an energy 
conversion factor (ECF) using a power law fit with the absorption column
density fixed to the Galactic value \citep[1.2$\times 10^{20}$
cm$^{-2}$, ][]{dic90}.
Only one ECF with
ECF=3.76$\times 10^{-11}$ ergs s$^{-1}$ cm$^{-2}$ count$^{-1}$
was applied for the whole light curve. 
As shown below in \S\ref{xrayspec}, there is
no obvious change of the spectral parameters between early and later
observations.

\section{\label{results} Results}

\subsection{Positions}

\begin{figure*}
\epsscale{1.5}
\plotone{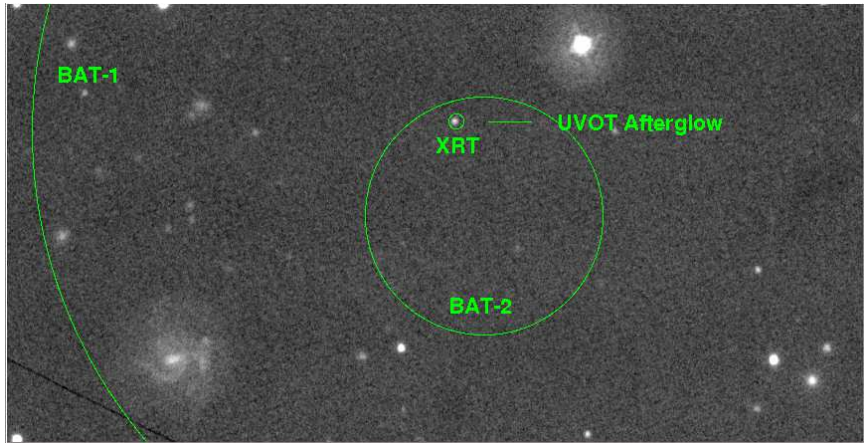}
\caption{\label{050603_error} \swift\ UVOT V-filter image 
of the field around GRB\,050603
with the BAT and XRT error circles superimposed. Positions are
given in Table~\ref{position}. BAT-1 refers to the onboard processed position
and error circle and BAT-2 to the error circle from the data processed on the
ground.
}
\end{figure*}

All positions given for GRB\,050603 are listed in Table~\ref{position}. 
Figure~\ref{050603_error} displays the UVOT V-filter
image of the GRB\,050603 field with
the BAT and XRT error circles superimposed as given in Table~\ref{position}. 
The XRT position given in  Table~\ref{position} is corrected for the XRT boresight
offset \citep{moretti05} and differs slightly
from the positions given by \citet{grupe05} and \citet{racusin05}.
This new XRT position
is in excellent agreement with
the UVOT, optical, and radio positions \citep{brown05a, berger05a, cameron05}.

\subsection{BAT}

\begin{figure*}
\epsscale{1.8}
\plottwo{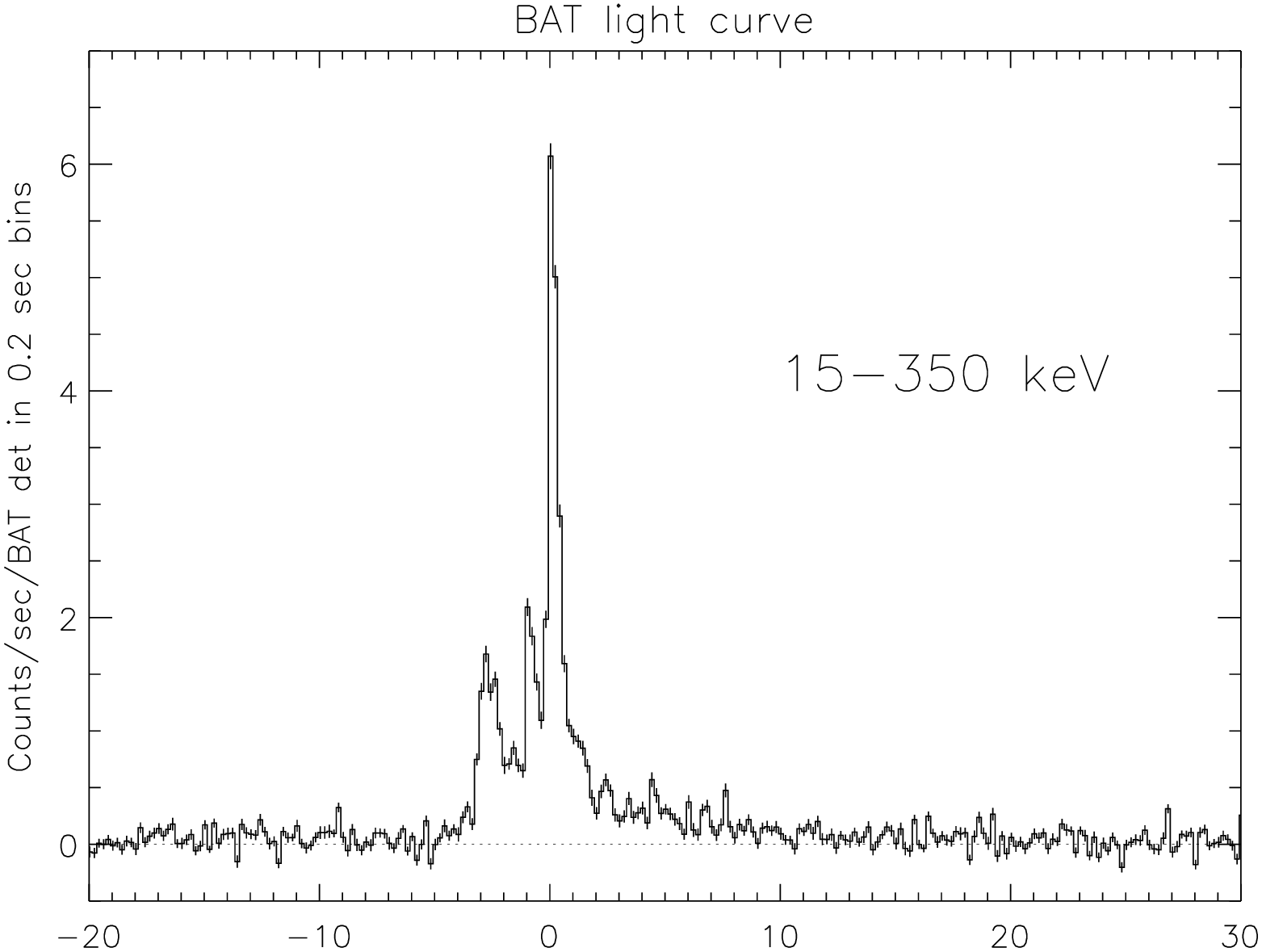}{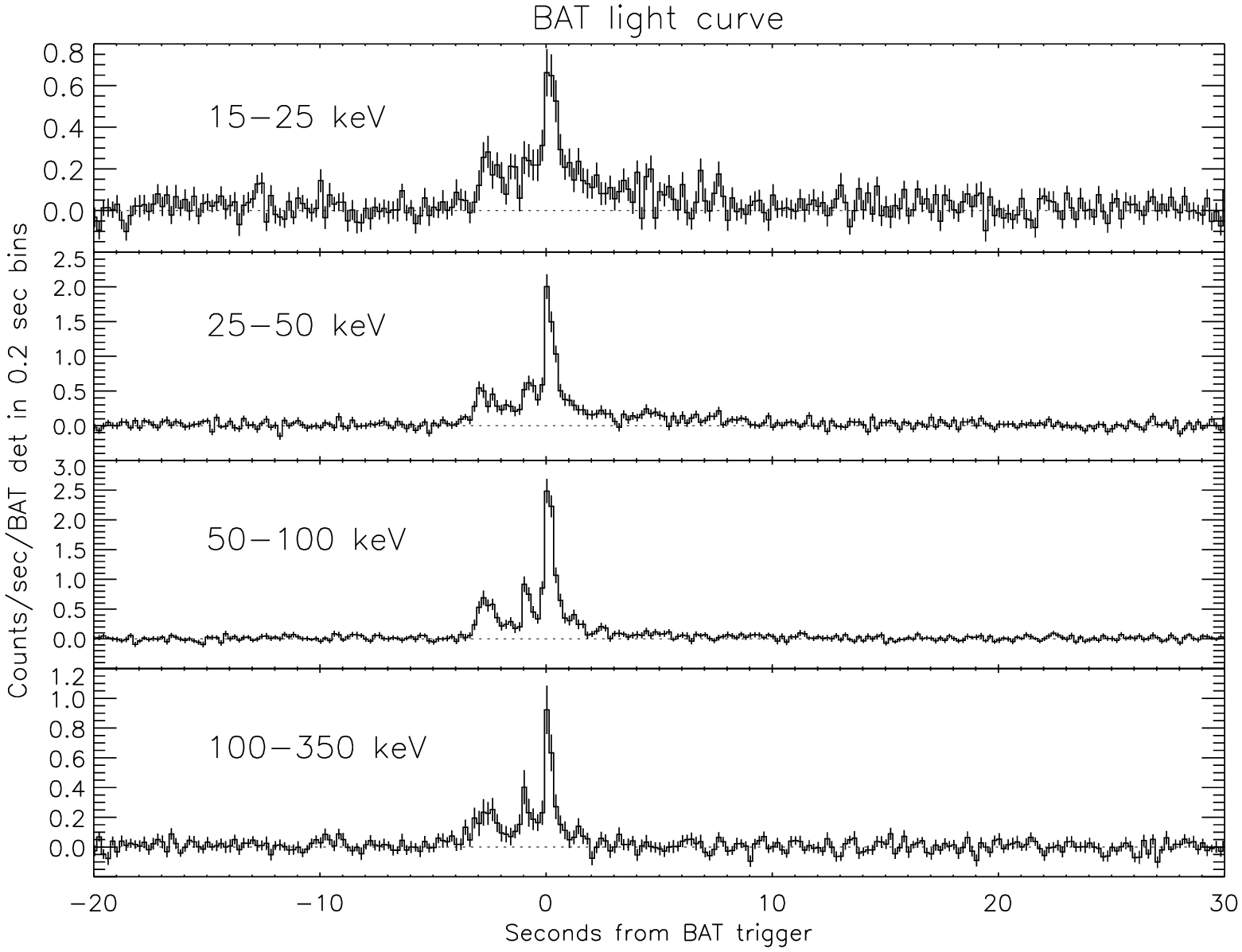}
\caption{\label{050603_bat_lc} \swift\ BAT light curves, left panel shows the
light curve in the energy range 15-350 keV, and the right panel the light curve 
split into 4 subranges (see text).}
\end{figure*}

\begin{figure*}
\epsscale{1.0}
\plotone{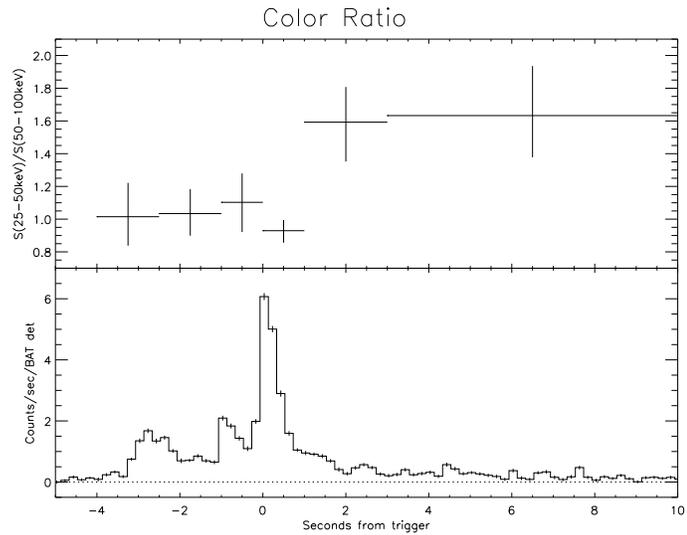}
\caption{\label{050603_bat_lc_color} \swift\ BAT color ratio (top) and light curve
(bottom).}
\end{figure*}

The BAT mask-weighted light curve (Figure~\ref{050603_bat_lc}) shows three
fast-rise-exponential-decay (FRED) like spikes with peaks at 2.7 and 0.85 s
before the trigger and 0.15 s afterwards. Each spike had a width of 0.6 s FWHM. 
The left panel of
Figure~\ref{050603_bat_lc} displays the light curve of the whole 15-350 keV
energy band. The right panel shows the light curve split into 
four energy bands: 15--25~keV, 25--50~keV, 50--100~keV, and 100--350~keV.   
The BAT light curve
shows a harder spectrum during the spikes compared to the BAT observation after
the spikes (Figure~\ref{050603_bat_lc_color}). 

\begin{figure*}
\epsscale{1.0}
\plotone{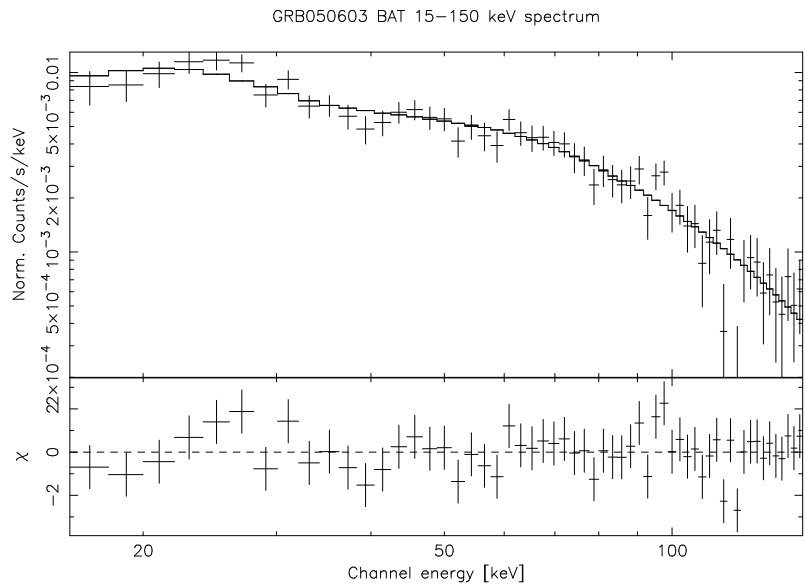}
\caption{\label{050603_bat_spec} \swift\ BAT 15-150 keV spectrum}
\end{figure*}

The time-averaged spectrum between $T_0$-3s and $T_0$+18s
in the 15-150 keV band shown in
Figure~\ref{050603_bat_spec} 
is well fitted by
a single power law with an energy spectral slope
$\beta_{\gamma}=0.17^{+0.07}_{-0.08}$ ($\chi^2/\nu$=51/57). The spectrum
is background subtracted. It has the standard 80-channel BAT
energy binning.

GRB\,050603 had $T_{90}$=12\plm2s which classifies it as a long burst.
The fluence in the observed
15--150~keV band was (7.6\plm0.3)$\times10^{-6}$ ergs
cm$^{-2}$. Based on the redshift z=2.812 \citep{berger05b},
the k-corrected rest-frame 100--500~keV and 20~keV--2~MeV 
total  isotropic equivalent energies
$E_{\rm \gamma, iso}$=3.2$\times10^{53}$
ergs and $E_{\rm \gamma, iso}$=1.26$\times10^{54}$ ergs, respectively,  
are some of
the largest measured among all GRBs detected by \swift\  \citep{nousek05}.
The total energies are comparable 
 to other high-energetic pre-\swift\ bursts such as
GRB\,990123 \citep{briggs99, corsi05}, GRB\,000131 \citep{andersen00}, 
or GRB\,010222 \citep{zand01}.

\subsection{XRT data}

\subsubsection{\label{xrt_lc} XRT light curve}

\begin{figure*}
\epsscale{1.5}
\plotone{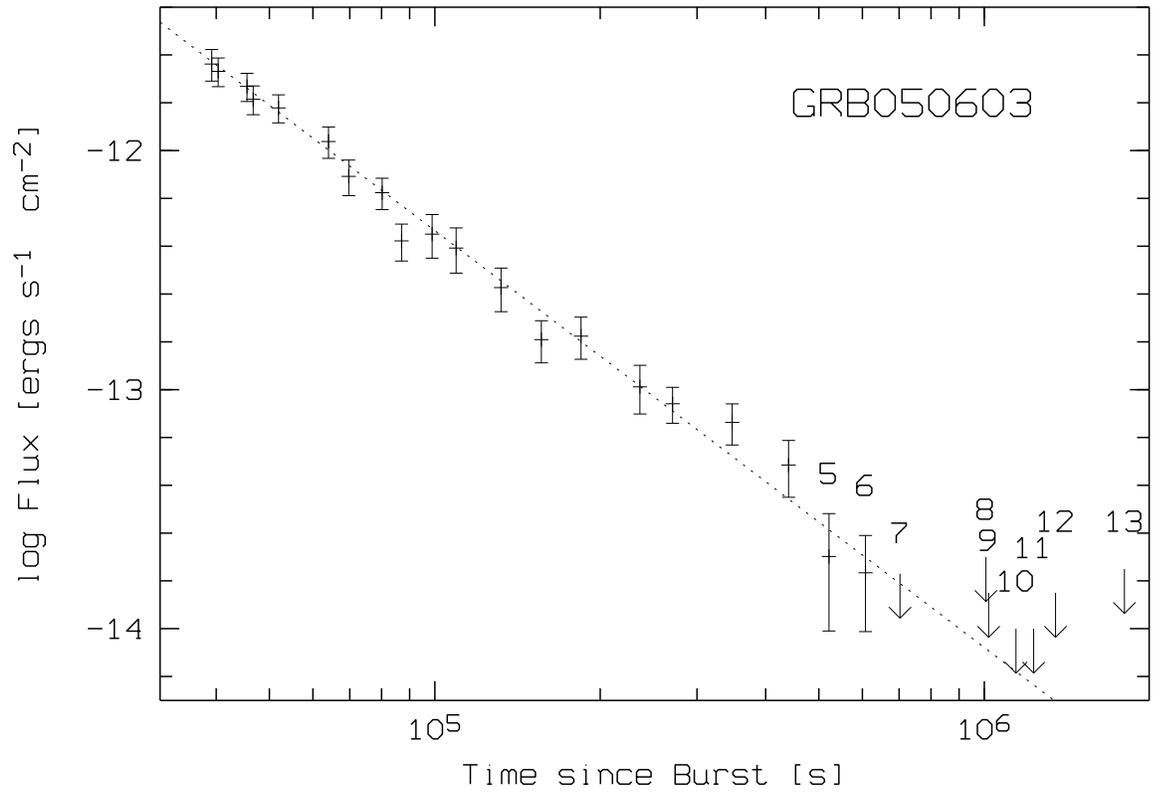}
\caption{\label{050603_lc} \swift\ XRT 0.3-10.0 keV unabsorbed flux
light curve of GRB\,050603 fitted in XSPEC by
a single power law with $\alpha_3$=1.76\plm0.07. 
The downwards arrows represent 3$\sigma$ upper limits. The Numbers 5-13
refer to the
segments as given in Table~\ref{obs_log}.
}
\end{figure*}

Figure~\ref{050603_lc} displays the 0.3-10 keV flux
X-ray light curve of GRB\,050603. The initial count rate at the beginning of the
XRT observation was 0.06 counts s$^{-1}$ which converts to a 0.3-10.0 keV
unabsorbed flux
$F_{\rm X}=3\times10^{-12}$ ergs s$^{-1}$ cm$^{-2}$. The 2.0-10.0 keV unabsorbed
flux was  $F_{\rm X}=2\times10^{-12}$ ergs s$^{-1}$ cm$^{-2}$. 
The count rate was low enough that the data are not affected by
pileup.
The decay slope, derived from the 0.3-10.0 keV flux light curve shown in
Figure~\ref{050603_lc},
is unusually steep with
$\alpha=1.76^{+0.15}_{-0.07}$. 
This makes
GRB\,050603 a relatively rapidly fading afterglow in its late phase
\citep{nousek05}.  The data are consistent with one simple
power law throughout the observation with a $\chi^2/\nu$=14.3/17. 

Just before the detection of GRB\,050603 the XRT MOS CCD was hit by a
micro-meteorite which caused severe damage to columns 294 and 320. 
These columns (and several adjacent ones) have been %turned off 
 disabled on-board % we can't actually ``turn off'' columns
and are not useable.  If the PSF of a source overlaps these bad columns, 
the measured flux can be incorrect.  Since the source position on the detector 
changes with each orbit, this can lead to errors in the light curve.  
We have verified that the GRB was not positioned on the bad columns 
during this observation.

\subsubsection{\label{xrayspec} X-ray spectral analysis}

\begin{figure*}
\epsscale{1.2}
\plotone{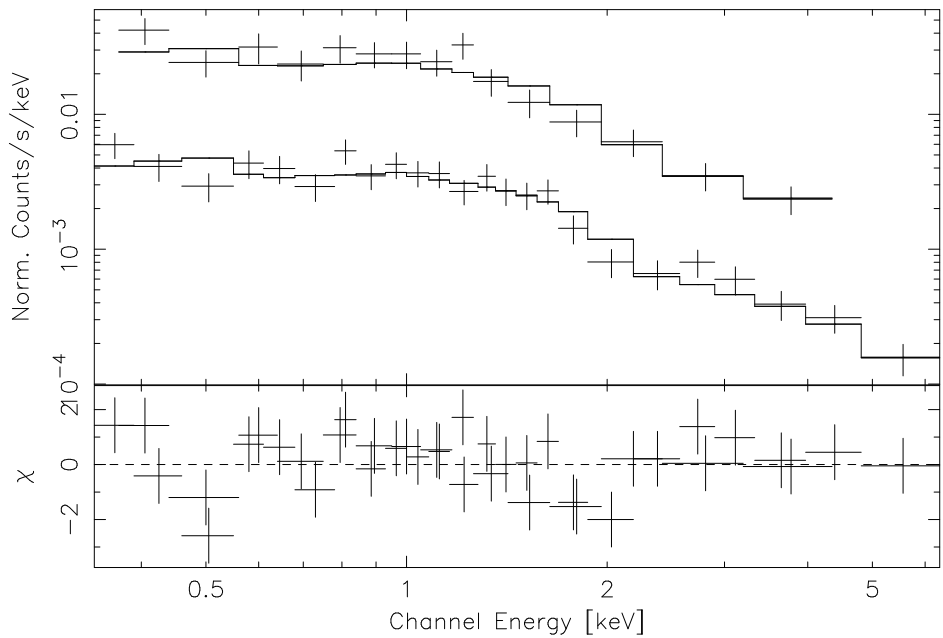}
\caption{\label{050603_spec} X-ray spectra of the observation segments 
001 (top) and 002 (Table\,\ref{obs_log} 
of GRB\,050603 fitted by a single power law (see Table~\ref{spectral}). The
residuals at 0.5 and 2 keV are caused by systematic 
errors in the auxiliary response file calibration (\S\ref{xrayspec}). 
}
\end{figure*}

Table~\ref{spectral} lists the results of the spectral analysis of the XRT
data. A simple power law model with Galactic absorption 
fits the spectra well for the segments 001 and 002 (Table\,\ref{obs_log})
data, resulting in
X-ray energy spectral slopes 
$\beta_{\rm X}$=0.80\plm0.17 and 0.62\plm0.13 for segments 001 and 002,
respectively.
The simultaneous fit to the segment 001 and 002 data is shown in
Figure~\ref{050603_spec}. The energy ranges of the two spectra are different due
to the S/N and teh binning of the channels using {\it grppha}.
This fits results in an X-ray spectra slope
$\beta_{\rm X}$=0.71\plm0.10. In all cases, no additional intrinsic absorption is
required. Leaving the absorption column density as a free parameter results
in an absorption column density which is significantly below the Galactic
value. 
As shown in Figure~\ref{050603_spec} there are some apparent residuals
 around 0.5 and 2.0 keV. These features are due to systematic errors in
the still ongoing calibration of the auxiliary response file 
\citep[see the XRT calibration document XRT-OAB-CAL-ARF-v3\footnote{
The calibration document XRT-OAB-CAL-ARF-v3 can be found under:
http://swift.gsfc.nasa.gov/docs/heasarc/caldb/swift/docs/xrt/index.html} 
and][]{romano05}.  

% we don't need to call this much attention to the residuals.

\subsection{UVOT}

\begin{figure*}
\epsscale{2.5}
\plotone{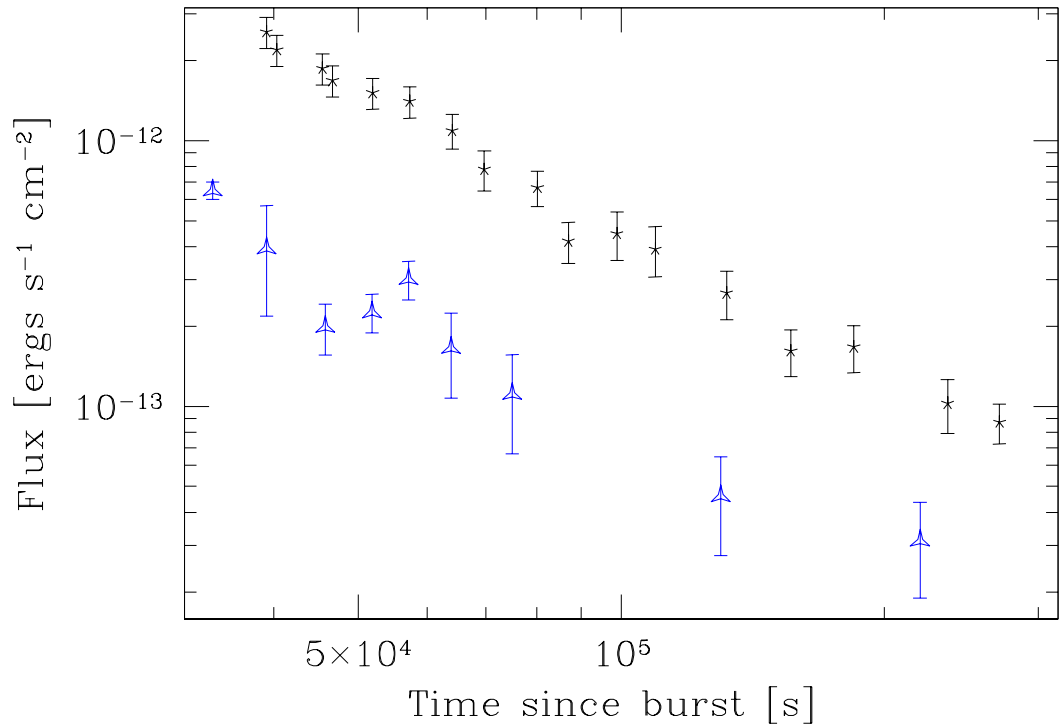}
\caption{\label{050603_lc_xrt_uvot} Combined XRT and UVOT V-filter light curves.
The UVOT V fluxes were multiplied by a factor of
10$^{6}$ in order to plot them  together
with the XRT data. The XRT data are displayed as crosses and the UVOT V-band
data as triangles.}
\end{figure*}

The V magnitudes for GRB\,050603 are listed in Table 4 and the V-band light curve is
shown in Figure~\ref{050603_lc_xrt_uvot}. 
 The first measurement, V=18.2 at 9.5 hours after the
burst, is brighter than any other optical \swift\ afterglow at the
same time, many of which are not optically detected at
all (see e.g. Roming et al. 2005b), and is among the brightest
of all optical afterglows with the
exception of GRB\,030329 \citep[see e.g. ][]{berger05c}.
 The X-ray and optical decay slopes are similar,
with the UVOT points decaying with a power law slope
of 1.8\plm0.2 with $\chi^2/\nu$=18/7.
There is no indication of a break
near 12 hours as suggested by Berger \& Becker (2005),
though the errors and wide temporal sampling do not
allow a definitive statement.  The steepness of the
decay indicates that the break may have already
occurred before our observations began, which would put a
  putative break at less than T+9 hours post-burst.

The large value of $\chi^2/\nu$ indicates that the
deviations from the simple power law may be real,
though the largest deviations are only 2 $\sigma$.  Li et
al. (2006), in an independent analysis of the UVOT
data, also concluded that this afterglow exhibited
real fluctuations from a powerlaw decay.  The power
law decay reported by Li et al. (2006) is 1.86\plm0.06,
which is consistent with our results and also in
agreement with that seen in the X-rays.

\subsection{Other wavelengths}

GRB\,050603 was the target of several ground-based observations at  
radio and optical wavelengths. \citet{cameron05} reported the Very Large
Array (VLA) position at 8.46 GHz as listed in Table~\ref{position}. The flux
density was 262\plm41 $\mu$Jy at 8.4 hours after the burst.
The afterglow was also observed by SCUBA at 450 and 850 $\mu$m at 11.2 and
13.3 hours after the burst, respectively.
\citet{barnard05} reported of a 
1.2$\sigma$ detection with a flux density of
2.408\plm1.973 mJy at 450 $\mu$m, but no detection at 850$\mu$m. 

\citet{berger05a} reported an $R$-band detection of the afterglow with the 2.5m
du Pont telescope at Las Campanas Observatory
3.4 hours after the burst with $R$=16.5 mag. The optical position is given in
Table~\ref{position}. \citet{berger05b} measured the redshift of the afterglow
as z=2.821 based on a single emission line which was interpreted as Ly$\alpha$.
This observation was performed with the Magellan/Baade telescope 2.13 days after
the burst. 

GRB\,050603 was also detected by the IBIS instrument on board
INTEGRAL in the 40-300 keV energy range \citep{gotz05}. However, because the
burst was outside the field of view it could not be localized. The duration of
the burst seen by IBIS agreed with the results from the BAT.
\citet{golenetskii05} reported the detection of GRB050603 by Konus-Wind in
the observed 20keV-3MeV range.
Its fluence in this energy band was $(3.41\pm0.06) \times 10^{-5}$ 
ergs cm$^{-2}$ with a duration of 6~s and an 
observed $E_{\rm peak}$=349\plm28 keV.

\section{\label{discuss}Discussion}
GRB\,050603 was a particularly  bright and energetic burst. The 15-150 keV fluence of 
7.6$\times~10^{-6}$ ergs cm$^{-2}$ places it among the top 10\% of \swift\ bursts. The 
isotropic energy  of 1.26$\times10^{54}$ ergs makes it one of the most energetic bursts
ever detected. Combined with the rest-frame $E_{\rm peak}$=1.3 MeV, GRB\,050603 is
consistent with the Amati relation
\citep{amati02}. 

With an optical magnitude of V=18.2 9.5 hours after the burst it was also the
brightest optical afterglow seen so far by \swift\ at this late time after the
burst. 
The only burst that had a comparable magnitude (V=18.9) at 10 hours after
the burst was GRB\,050525 \citep{blustin06}. All other bursts were far below this
magnitude at this time after the burst. In X-rays, however, the afterglow of 
GRB\,050603 was not the brightest one among other \swift\ bursts observed at similar
times after the trigger.  

The X-ray afterglow of GRB\,050603 decayed with a rather steep slope of
$\alpha$=1.76\plm0.07, compared with the mean decay slope of
$\alpha=1.34 \pm 0.32$ for 27 bursts
listed in \citet{nousek05}. 
This is intermediate between the slopes of $\sim 1.3 \pm 0.1$ obtained from
Chandra grating observations, XMM
observations, and Beppo-SAX afterglows discussed in \citet{gendre06}, but is
somewhat shallower than the mean decay slope of \ax=2.0\plm0.3 from their
survey of Chandra imaging afterglow observations.
It is also intermediate between the expectations for the ``normal''
afterglow slope following the end of energy injection by the central
engine, and the steeper slope expected following the end of energy
injection by the central engine.
We note, however, that the decay slope in the V-band filter ($\alpha=1.8\pm0.2$) is consistent with the X-ray
decay slope, consistent with the typical signature of an afterglow after the jet break
\citep{sari99}.
This suggests the possibility that  GRB\,050603 had an early
jet break before 2.9 hours (rest-frame) after the burst. 
If we assume that the jet break happened before the start of the
UVOT and XRT observations, we can use the dependence of jet angle on
break time \citep{sari99, frail01} to place a limit on the jet opening angle of 
$\Theta_j < 1\degdot3$, where we have used
$E_{\rm  iso}=1.26\times 10^{54}$ ergs and we assumed
the density of the circum-burst matter is 0.1 cm$^{-3}$.
We note that \citet{bloom03} argue for a larger typical circum-burst
density of 10 cm$^{-3}$, which would imply a slightly larger jet angle
limit of 2\degdot3.

In order to check the jet interpretation, we compare the observed
\ax=$1.76^{+0.15}_{-0.07}$ and \bx=0.71\plm0.10 to the prediction of
the jet model. 
For the
familiar case of $p>2$ (where $p$ is the electron spectral index),
one requires that $\alpha_{\rm X} - 2\beta_{\rm X} = 0$ ($\nu > \nu_c$) or
$\alpha_{\rm X} - 2\beta_{\rm X} = 1$ ($\nu_m < \nu <\nu_c$), where $\nu_m$ is the
typical synchrotron frequency (injection frequency) and $\nu_c$
is the cooling frequency \citep{rhoads99}. 
In this case, we have $\alpha_{\rm X} - 2\beta_{\rm X} = 0.34^{+0.18}_{-0.12}$.
Both cases are inconsistent with the data at the $>2.8\sigma$ level. 
We therefore consider the case 
of $1 < p <2$ \citep{dai01}. 
For $\nu > \nu_c$, this case has $2\alpha_{\rm X}-\beta_{\rm X}=3$,
while the data have $2\alpha_{\rm X}-\beta_{\rm
  X}=2.81^{+0.32}_{-0.12}$, in good agreement with theory.
In this regime the electron index is $p=2\beta_{\rm X} = 1.4 \pm 0.2$.
This case also has $\alpha_{\rm O}=\alpha_{\rm X}$ as long as the
self-absorption frequency is below the optical band, in agreement with
the observations.  In this case, we expect $\beta_{\rm O} = \beta_{\rm
  X}$ if the optical band is above the cooling frequency, or
$\beta_{\rm O}=0.2 \pm 0.1$ if the optical band is between the
injection frequency and the cooling frequency; however, 
with only V-band data we
cannot constrain $\beta_{\rm O}$ directly.  The observed spectral index
between the X-ray and optical bands is $\beta_{\rm OX}$\footnote{Here we define
the optical to X-ray spectral slope $\beta_{\rm OX}= log (5460\AA \times f_{\rm
5460\AA}) - log (1{\rm keV} \times f_{\rm 1KeV}$)}=0.04.
These spectral slopes suggest that there is
a break in the spectrum between the optical and X-rays suggesting
that the optical band
is between the injection and cooling frequencies. 
We therefore conclude that the jet interpretation
fits these data provided there is a flat electron energy spectrum $p=1.4$
\citep{dai01}. We note however, that this interpretation is at odds with the
expectations from \citet{liang05} who find the relation 
$E_{\gamma,iso,52}=0.85\times(\frac{E_{\rm peak}}{\rm
100~keV})^{1.94}\times t^{-1.34}$, which predicts a jet break in the optical light
curve at 1 day after the burst in the rest-frame, or 3.8 days in the observed frame.

\acknowledgments

We would like to thank the anonymous referee for valuable comments and
suggestions that improved the paper.
We would also like to thank Vicki Barnard and Brian Cameron 
for their information on the SCUBA and VLA 
observations. 
We made use of data obtained through the High Energy 
Astrophysics Science Archive Research Center Online Service, provided by the 
NASA/Goddard Space Flight Center.
This work has been supported by NASA contract NAS5-00136, SAO grant
GO5-6076 BASIC and NASA grant NNG05GF43G.



\begin{deluxetable}{lccrl}
%\tabletypesize{\scriptsize}
\tablecaption{Log of of the \swift\ XRT observations of GRB\,050603
\label{obs_log}}
\tablewidth{0pt}
\tablehead{
\colhead{Segment} & \colhead{T-start\tablenotemark{1}} & 
\colhead{T-stop\tablenotemark{1}} &
\colhead{$\rm T_{exp}$\tablenotemark{2}} & 
\colhead{Binning\tablenotemark{3}}  
} 
\startdata
001 & 2005-06-03 15:49:47 & 2005-06-03 22:32:14 &  7122 & 50 \\
002 & 2005-06-04 00:02:16 & 2005-06-06 21:10:00 & 72838 & 50 \\
003 & 2005-06-07 00:04:50 & 2005-06-07 23:15:57 & 14612 & 25 \\
004 & 2005-06-08 00:26:48 & 2005-06-08 23:22:58 & 11858 & 15 \\
005 & 2005-06-09 00:32:42 & 2005-06-09 23:28:58 &  9840 & 10 \\
006 & 2005-06-10 00:39:43 & 2005-06-10 23:35:57 & 11184 & 10 \\
007 & 2005-06-11 00:55:38 & 2005-06-11 22:05:57 &  6463 & ul\tablenotemark{4} \\
008 & 2005-06-14 20:23:42 & 2005-06-14 23:59:56 &  3766 & ul\tablenotemark{4} \\
009 & 2005-06-15 00:58:31 & 2005-06-15 23:59:59 & 16453 & ul\tablenotemark{4} \\
010 & 2005-06-16 00:04:59 & 2005-06-16 23:59:57 & 11475 & ul\tablenotemark{4} \\
011 & 2005-06-17 01:18:43 & 2005-06-17 06:39:57 &  6292 & ul\tablenotemark{4} \\
012 & 2005-06-18 00:05:01 & 2005-06-20 23:03:20 & 44626 & ul\tablenotemark{4} \\
013 & 2005-06-24 00:28:17 & 2005-06-24 02:35:53 &  3284 & ul\tablenotemark{4}  
\enddata

\tablenotetext{1}{Start and End times are given in UT}
\tablenotetext{2}{Observing time given in s}
\tablenotetext{3}{Number of photons per bin in the light curve}
\tablenotetext{4}{The source is not detected and only an upper limit
can be given.}
\end{deluxetable}

\begin{deluxetable}{lllrl}
%\tabletypesize{\scriptsize}
\tablecaption{Calculated positions of GRB050603. 
BAT1 and BAT2 
refer to the positions as shown in Figure~\ref{050603_error}. 
\label{position}}
\tablewidth{0pt}
\tablehead{
\colhead{Position from} & \colhead{RA (J2000)} & 
\colhead{Dec (J2000)} &
\colhead{Position error\tablenotemark{1}} &  
\colhead{reference}
} 
\startdata
BAT on-board (BAT1) & 02 39 55 & --25 10 57 & 240 & \citet{retter05} \\
BAT ground (BAT2) & 02 39 56 & --25 11 41 & 60 &  \citet{fenimore05} \\
XRT  & 02 39 56.73 & --25 10 54.36 & 3.9 &  \citet{moretti05} \\
UVOT V-filter & 02 39 56.8 & --25 10 54.9 & 1.0 & \citet{brown05a} \\
Las Campanas R-band & 02 39 57 & --25 10 54 & 0.5 & \citet{berger05a} \\
VLA 8.5 GHz & 02 39 56.9 & --25 10 54.6 & 0.1 & \citet{cameron05} 
\enddata

\tablenotetext{1}{Position uncertainty is given in arc seconds}

\end{deluxetable}

\begin{deluxetable}{lcc}
%\tabletypesize{\scriptsize}
\tablecaption{Spectral analysis of GRB\,050603 with a power law fit 
 with the column density fixed at the Galactic value \citep[$1.19 \times 10^{20}$
cm$^{-2}$, ][]{dic90} 
\label{spectral}}
\tablewidth{0pt}
\tablehead{
\colhead{Segment} & 
\colhead{$\beta_{\rm X}$} & 
\colhead{$\chi^2/\nu$} 
} 
\startdata
001 &  0.80\plm0.17 &  13.5/13 \\
002 &  0.62\plm0.13 &  24.3/22 \\
001 + 002  & 0.71\plm0.10 &  39.2/36 \\
\enddata

\end{deluxetable}

\begin{deluxetable}{rrcc}
%\tabletypesize{\scriptsize}
\tablecaption{V-magnitudes GRB\,050603 from the UVOT observations 
\label{uvot_mag}}
\tablewidth{0pt}
\tablehead{
\colhead{$T_{\rm after~burst}$\tablenotemark{1}} 
& \colhead{$T_{\rm exp}$\tablenotemark{2}} &
\colhead{V mag} & 
\colhead{V error} 
} 
\startdata
 34093  &  1298	&   18.19 &  0.08 \\
 39276  &  110	&   18.73 &  0.40 \\
 45864  &  1772	&   19.47 &  0.21 \\
 51840  &  2104	&   19.33 &  0.17 \\
 57096  &  1200	&   19.02 &  0.17 \\
 63900  &  1205	&   19.67 &  0.33 \\
 75024  &  2056	&   20.1  &  0.37 \\
129996  &  10458  &  21.06 &  0.37 \\
219708  &  10945  &  21.48 &  0.36 \\
\enddata

\tablenotetext{1}{The times after the burst are given in s and mark the middle
of the time bin.}
\tablenotetext{2}{Exposure times $T_{\rm exp}$ given in s.}.

\end{deluxetable}

\end{document}